\documentclass[twocolumn,letterpaper,secnumarabic,amssymb, nobibnotes, aps, prl]{revtex4-1}        
\usepackage{epsfig}
\usepackage{dcolumn}
\usepackage{epstopdf}
\usepackage{graphicx}
\begin{document}

\title{Adiabatic Variations of Quantum Heat of a quantum body}  

 \author{S. Selenu}
\affiliation{}

\begin{abstract}
\noindent In this article it will be introduced a new theorem, can be considered a generalization of Hellmann-Feynman theorem\cite{Feynman}. The latter used in conjunction with the quantization of the free energy\cite{SelenuEntropy} of a quantum system allows to derive strightly the electronic Heat variations  of a quantum electronic system, in its condensed phase of  eigenstates, showing its agreement with classical thermodynamics, making then possible to desing ab inito quantum  models of the electronic structure in its thermodynamical states.
\end{abstract}

\date{\today} 
\maketitle

 \section{Introduction}

\noindent  
Since the advent of classical thermodynamics \cite{Fermi} and classical mechanical statistics\cite{11} the introduction of state functions such as the Entropy allowed to calculate or either quantify physical properties of matter in its thermodynamical states where an effort have been put on the making rational of  physical phenomena involved from a corpuscolar point of view. The advent of quantum mechanics put on clear evidence the effect it could have the quantum wave nature of matter particles, as it could be thougth being  the electronic field of a quantum body,  on the thermodynamical properties of the latter. The introduction in \cite{SelenuEntropy} of an Entropy functional of the quantum probability of an event happening in a DFT context made possible determine new quantum state functions of the system while recognizing them reducing to observables  encountered in classical thermodynamics, i.e. Entropy, Enthalpy and Gibbs free Energy\cite{SelenuEntropy}. Nonetheless it will be introduced here in this article a new theorem making us able evaluating Heat  variations of a quantum body in its condesed phase reaching  an agreement with macroscopics of physical systems as found also in classical thermodynamics. A built more on quantum field theory shows the agreement, from a Copenaghen point of view\cite{Heisenberg}, existing between the quantum expectation value of an observable and their macroscopic measurements routinely made on laboratories as their counterpart. In the next section it will be introduced the new thermodynamical model in order to  calculate heat variations on an adiabatic thermodynamical loop  recovering also the equipartition of the energy. Variations of the Entropy of an isolated system in thermal equilibrium will be also reported leaving conclusions at the end of the article.

\section{Entropy variations}
In this section it will be introduced a new theorem including Hellmann-Feynman theorem\cite{Feynman,Selenucurrent} showing how to calculate quantum mechanically the expectation value of an operator $\hat{O}$ smoothly varying with respect a parameter $\lambda$. The latter theorem will be employed in order to calculate the announced heat  variations, where an electronic charge density associated  to a quantum body is considered, and  given by the following relation $\rho=\sum_n f_n \Psi^*_n\Psi_n$ times the electronic charge, i.e. a product of electronic elementary charge $e$ and the quantum probability $\rho$ of an event happening, calculated as the weighted product of electronic  wave functions of the quantum system at a state $n$ with occupation numbers $f_n$ of the quantum eigenstate being filled. The latter probability density can be directly used in order to perform calculations of the potential energies usually encountered in quantum DFT, as s also the quantum free energy and Enthalpy, considered then eligible candidates for a further study of the quantum thermodynamics of  a condensed matter system at the thermal equilibrium. It will be introduced the theorem by firstly writing the expectation value of a smooth operator $\hat{O}$   with respect to eigenstates $\Psi_n$ of the system:

\begin{eqnarray}
\label{H1}
O=\sum_n f_n \langle \Psi_n |\hat{O}|\Psi_n\rangle
\end{eqnarray}

and let consider derivatives of $O_n=\langle \Psi_n |\hat{O}|\Psi_n\rangle$ with respect to a parameter $\lambda$ making varying its states parametrically along a physical transformation bringing the quantum body from an inital state $1$ to a final prescribed state $2$:

\begin{eqnarray}
\label{H11}
\frac{\partial O_n}{\partial_\lambda}=  \frac{\partial \langle \Psi_n |\hat{O}|\Psi_n\rangle}{\partial_\lambda}
\end{eqnarray}

took into account that,

\begin{eqnarray}
\label{H0}
\partial_\lambda \Psi_n= \frac{1}{N'}\langle \Psi_n |\partial_\lambda| \Psi_n \rangle \Psi_n
\end{eqnarray}
being $N'$ the normalization constant of the eigen function $\Psi_n$  in its state $n$, it is possible derive  eq.(\ref{H11}) with respect to $\lambda$ obtaining the following relation:

\begin{eqnarray}
\label{H110}
\frac{\partial O_n}{\partial_\lambda}=  \langle\partial_\lambda \Psi_n |\hat{O}|\Psi_n\rangle + \langle\Psi_n |\partial_\lambda \hat{O}|\Psi_n\rangle+  \langle \Psi_n |\hat{O}|\partial_\lambda\Psi_n\rangle
\end{eqnarray}

that  can be recasted as:
\begin{eqnarray}
\label{H01}
\frac{\partial O_n}{\partial_\lambda}&&= \frac{1}{N'} \langle\partial_\lambda \Psi_n |\Psi_n \rangle  \langle \Psi_n |\hat{O}|\Psi_n\rangle + \langle\Psi_n |\partial_\lambda \hat{O}|\Psi_n\rangle \\\nonumber &&+  \frac{1}{N'} \langle \Psi_n |\hat{O}|\Psi_n \rangle \langle \Psi_n | \partial_\lambda\Psi_n\rangle
\end{eqnarray}

by direct substitution of eq.(\ref{H0}) in eq.(\ref{H110}). Last step of our demonstration consists of  recognizing the identity:
\begin{eqnarray}
\label{H00}
\langle\partial_\lambda \Psi_n | \Psi_n \rangle =-\langle \Psi_n |\partial_\lambda\Psi_n \rangle 
\end{eqnarray}
due to normalization conditions on electronic waves, it making possible recast  eq.(\ref{H01}):
\begin{eqnarray}
\label{H2}
\frac{\partial O_n}{\partial_\lambda}&&= \langle\Psi_n |\partial_\lambda \hat{O}|\Psi_n\rangle 
 \end{eqnarray}

Above result allows deriving the expectation value $O$  of eq.(\ref{H1}) directly as:

\begin{eqnarray}
\label{H1}
\frac{\partial O}{\partial_\lambda}=\sum_n f_n \langle \Psi_n |\partial_\lambda \hat{O}|\Psi_n\rangle
\end{eqnarray}

By considering the general expression of the quantum Entropy derived in \cite{SelenuEntropy}:

\begin{eqnarray}
\label{H20}
S=-k_b\sum_n f_n \langle \Psi_n |ln \rho|\Psi_n\rangle
\end{eqnarray}

Entropy variations of the quantum state of matter can be evaluated as:

\begin{eqnarray}
\label{H200}
{\partial_\lambda S}=-k_b\sum_n f_n \langle \Psi_n |\frac{1}{\rho}\partial_\lambda \rho|\Psi_n\rangle
\end{eqnarray}

Because of the linearity of the integral operator we can simplify the factor $\frac{1}{\rho}$,  summing eigenfunctions $\Psi_n$ in the integrals with respect to occupation numbers $f_n$ leaving to a simple formula:

\begin{eqnarray}
\label{H3}
{\partial_\lambda S}= -k_b\int \partial_\lambda \rho
\end{eqnarray}

that can be further written in a more convenient form:

\begin{eqnarray}
\label{H4}
{\partial_\lambda S}= -k_b \partial_\lambda N
\end{eqnarray}
being $N=\int \rho$ of the unormalized distribution function $\rho$\cite{SelenuEntropy}. Considering then a system having a  distribution of $N$ electrons,  depending on $\lambda$  and  in  their thermal equilibrium, by integration of eq.(\ref{H4})  with respect to $\lambda$ we reach the satisfactory result  already found naturally on  macroscopic thermodynamical  systems:

\begin{eqnarray}
\label{H6}
\Delta S= -k_b \Delta N
\end{eqnarray}

making possible calculate the quantum electronic heat delivered or either acquired along the thermo dynamical phase change of eigenstates equal to:

\begin{eqnarray}
\label{H7}
&&\Delta Q=-T\Delta S \\\nonumber 
&&\Delta Q= \Delta Nk_bT
\end{eqnarray}

being the latter the result admissible to the equipartition energy\cite{2} when is the case of a set  of filled eigen states, with two electrons per state,  of a spin unpolarized   inhomogenous electron gas. By considering also infinitesimal variations of the quantum Heat it is straight forward deriving, by its very definition, the following result:

\begin{eqnarray}
\label{H8}
\oint dQ=0
\end{eqnarray}

in close agreement with the first principle of thermodynamics\cite{Fermi} in adiabatic processes. Also when is the case of thermodynamical state transformation depending on a parameter $\lambda$ in isolated systems the Entropy of eq.(\ref{H6}) stays unaltered,  due to the conservation of electronic charge in the volume of the system, not being the latter affected by any exchange of heat or either mass the electrons may transport on a form of an electronic current varying the number of states in the volume of the body, reaching then:

\begin{eqnarray}
\label{H5}
{\Delta S}= 0
\end{eqnarray}

It is clear that in thermodinamical processes on isolated systems  the entropy of the system does not vary with a null exchanging of quantum heat $\Delta Q=0$ in agreement with eq.(\ref{H5}) and the first principle of thermodynamics. In the next section it will be reported conclusions of the article. 

\section{Conclusions}
In this paper it has been shown a strict consequence of the definition of Entropy reported in\cite{SelenuEntropy} can be expressed as the following statement:"In thermally adiabatic processes  the quantum variations of the electronic heat  is zero on a thermodynamic cycle". This statement put us on the position of searching for a generalization of this result to any thermo dynamical process in a future work. On turn a generalization of the Hellmann-Feynman\cite{Feynman} theorem is reported showing how to calculate any quantum expectation value derivatives.

\end{document}